\documentclass{JHEP}

\newcommand{\eea}{\end{eqnarray}}
\newcommand{\bea}{\begin{eqnarray}}
\newcommand{\be}{\begin{equation}}
\newcommand{\ee}{\end{equation}}

\newcommand{\lan}{\langle}
\newcommand{\ran}{\rangle}
\usepackage{epsfig}
\usepackage{amssymb}

\title{\huge \boldmath $\widehat g$-coupling ($g_{_{B^*B\pi}}$, $g_{_{D^*D\pi}}$)\\
  A quark  model with Dirac equation}
\author{\large Damir~Becirevic$^{\mathrm a}$,~~Alain~Le Yaouanc$^{\mathrm b}$\\
\hfill \\
$^{\mathrm a}$ Dip. di Fisica, Universqit\`a ``La Sapienza" 
and INFN, Sezione di Roma,\\ 
P.le A. Moro, I-00185 Rome, Italy \\
\vspace*{2.5mm}
$^{\mathrm b}$ Laboratoire de Physique Th\'eorique~\footnote{Laboratoire 
associ\'e au Centre National de la Recherche
Scientifique - URA D00063}\ , B\^atiment 210,  
Universit\'e de Paris XI, \\
91405 Orsay Cedex, France}

\abstract{We discuss the strong coupling of heavy mesons to a pion  
$\widehat g$, in the heavy quark limit. This quantity is quite remarkable 
since its values as estimated by different methods (various quark models 
and the QCD sum rules), are surprisingly different. The present quark models 
are mostly based on free spinors and their predictions depend crucially
on the choice of the light quark mass. We propose
a quark model based on the Dirac equation in a central potential, 
which gives a more 
refined description of Dirac spinors. We show that within such a Dirac 
model, the value of $\widehat g$ is stable and large: $\widehat g = 0.6(1)$, where we assume no quark current renormalization ($(g_A)_q = 1$). Such 
a large result is strongly constrained by requiring that the model 
parameters fit the spectrum; we show that this implies a large ``effective''
 light mass. It is also supported phenomenologically by a similar situation 
 with heavy baryons, as well as by experience with nucleon (if one 
 invokes additivity). We also calculate the couplings to heavy meson excitations, and show that the Adler--Weisberger sum rule is well 
 saturated by a few levels (in contrast to the case of small $\widehat g$). 
 We discuss uncertainties of our approach, and rise several questions which remain to be answered. The main mystery is the large, unusual discrepancy 
 with QCD sum rules for $\widehat g$, whereas a good agreement is 
 found for orbital excitations.}


\preprint{LPT-TH 99/04\\
ROME 99/1239}

\begin{document}

\section{Introduction}
\label{intro}
A precise value of the coupling of pion to heavy-light mesons 
($g_{B^*B\pi}$, $g_{D^*D\pi}$, or $\widehat g$  which is (in the heavy quark limit) related to $g_{B^*B\pi}$ as, {$g_{B^*B\pi}\ \simeq \  ( 2 m_B / f_{\pi})\ \widehat g$}), has been often needed in phenomenology during the last decade~(see Ref.~\cite{Casalbuoni}, for review). $g_{B^*B\pi}$ is particularly important, since it can be helpful 
in the analysis of the semileptonic $B\to \pi$ decay. 
Namely, the exclusive modes, $B\to \pi(\rho) \ell \nu_{\ell}$, represent the most promising way to extract the value of the least known CKM matrix element, $\vert V_{ub}\vert$. For that, 
the essential theoretical input are form factors, describing 
the hadronic matrix elements in the {\bf whole} physical region. 
The calculations which incorporate QCD in their first principles 
(lattice QCD and QCD sum rules) cannot fully solve the problem, {\it i.e.} they do not cover the whole physical range of $q^2$, in particular the very small recoil region, near $q_{max}^2$. 
A simple complementary theoretical idea is the one of the 
nearest pole (a vector meson) dominance (VMD). According to the extensive study in Ref.~\cite{Burdman}, in the heavy quark limit, the form factor $f^+_{B\to \pi}(q^2)$ is actually dominated by $B^*$, when close to $q^2_{max}$. Then, one has
\be
f^+(q^2)\ \simeq \ { 1\over 2\ m_{B^*}}\ 
{\quad f_{B^*}\ g_{B^*B\pi}\quad \over  1 \ -\ \frac{\textstyle{\large q^2}}{ \textstyle{\large m_{B^*}^2}} }  \; ,
\ee
where $f_{B^*}\ g_{B^*B\pi} m_{B^*}/2$, represents the residue of the form factor at that particular (nearest) pole\footnote{We take $f_\pi = 131~{\rm MeV}$.}. It includes $g_{B^*B\pi}$, 
whence a main interest in evaluating this quantity. 
Of course, the complete functional dependence for this form factor is different from VMD, {\it i.e.} for lower $q^2$ (larger recoils, $\vec q^2$), where the contributions of higher excitations become substantial, it becomes : 
\bea
&&\hfill \cr
f^+(q^2)&=& {1 \over 2}\ \left(\; \frac{f_{B^*}\,g_{B^*B\pi}\, m_{B^*}\;}{m_{B^*}^2 \ -\  q^2} \, +\, { \sum_{n}}\  \frac{\; f_{B_n^*}\,g_{B_n^* B\pi}\,m_{B_n^*}\;}{m_{B_n^*}^2\ -\ q^2}\right) \; ,\\
&& \nonumber
\eea
where $f_{B_n^*}\ g_{B_n^* B\pi}\ m_{B_n^*}$, are the $B\pi$ couplings to the the higher excitations, $B_n^*$. For the complete treatment, one should also 
account for multiparticle states, which cannot be incorporated in our simple approach.

There cannot be a direct experimental indication on $g_{B^*B\pi}$, because there is no phase space for the decay $B^* \to B \pi$. The available experimental results 
for $D^* \to D \pi$~\cite{PDG}~\footnote{$\Gamma(D^{*+})^{(exp.)} \times {\cal{B}}(D^{*+} \to D^0 \pi^+)^{(exp.)} < 90(2)\, keV$, amounts to the bound $\widehat g^2 < 0.57$.}, can be related to $g_{B^*B\pi}$, through the heavy quark symmetry ($HQS$). 
Still, the upper limit that can be obtained in this way is not very constraining (see Tab.~\ref{tab0}). One must then turn to theory.\\

A first lattice approach related to the above considerations would consist in comparing the lattice 
form factor data near zero recoil to VMD in the case of $D\to \pi$ (which is directly accessible on the lattice), and estimate $g_{D^*D\pi}$ and thus $g_{B^*B\pi}$. 
In this way, using the APE data~\cite{APE} for $f^+_{D(\vec p=0)\to \pi(\vec p_{lat})}(q^2)$, where $\vec p_{lat}$ is the smallest momentum that can be given to a particle on the lattice, one obtains $\widehat g=0.41(6)$. It is important to mention that the appropriate chiral extrapolation {\it was} performed. Results have also been obtained by working on $B \to \pi$ lattice data, which are obtained by extrapolation in the heavy quark mass~\cite{Lellouch}.

Very recently, a direct preliminary determination of $g_{B^*B\pi}$ on the lattice has been attempted by UKQCD~\cite{UKQCD}. The value they reported, $\widehat g=0.42(4)(8)$, is compatible with the one extracted from the form factor close to zero recoil.

We turn now to the other systematic theoretical methods for calculating the couplings: QCD sum rules~(QCDSR) and quark models~(QM). A new problem then arises: they lead to very different results for a seemingly simple quantity.
One would like to understand why. The main result we present here, concerns the coupling $\widehat g$. A short survey of the literature indicates that the value of $\widehat g$ is far from being established~({\it see} Tab.~\ref{tab0}). The overall conclusion drawn from different QCDSR calculations is that the coupling $g_{B^*B\pi}$ is small. In the heavy quark limit, their typical values are in the range $\widehat g \ \simeq \ 0.15 \ \div\  0.35$. On the other hand, the values proposed by different quark models are quite dispersed, ranging from $\widehat g = 1$ in nonrelativistic QM~\cite{Yan}, down to  $\widehat g = 1/3$, in certain partially relativistic models (see Ref.~\cite{Colangelo2}, for instance). Such discrepancies are very unusual, especially for ground state properties.

{\openup 4pt
\begin{table}[h!!!]
\begin{center}
\begin{tabular}{|c c c|} 
\hline
\multicolumn{3}{|c|}{\sf Lattice QCD}   \\
\hline
\phantom{\Huge{l}} \raisebox{.15cm}{\phantom{\Huge{j}}}
Ref.~\cite{UKQCD} \quad & \quad 0.42(4)(8) \quad & \parbox{5.3cm}{\footnotesize \sl Direct method by studying\\ the $\langle B\vert A_\mu(0)\vert B^*\rangle$.} \\
\vspace*{-4mm}
&& \\
\hline \hline
\multicolumn{3}{|c|}{\sf QCD Sum Rules}   \\
\hline
\phantom{\Huge{l}} \raisebox{.2cm}{\phantom{\Huge{j}}}
Ref.~\cite{Nardulli} &  \quad \parbox{1.6cm}{0.39(16)\\ 0.21(6)$^a$}$\biggr\}$ & \parbox{5.3cm}{\footnotesize \sl SR + soft pion limit; \\  $^{a)}$ Included ${\cal{O}}(\alpha_s)$ corrections.} \\
\phantom{\Huge{l}} \raisebox{.1cm}{\phantom{\Huge{j}}}
Ref.~\cite{Narison}& 0.15(3) & \parbox{5.3cm}{\footnotesize \sl Double moment SR.} \\
\phantom{\Huge{l}} \raisebox{.1cm}{\phantom{\Huge{j}}}
Ref.~\cite{Grozin}& \, 0.2 $\div$ 0.4 & \parbox{5.3cm}{\footnotesize \sl SR in external axial field.} \\
\phantom{\Huge{j}} \raisebox{.1cm}{\phantom{\Huge{l}}}
Ref.~\cite{Belyaev}& 0.32(2) & \parbox{5.3cm}{\footnotesize \sl Light cone SR (calcul. of {$g_{B^*B\pi}$}).} \\
\phantom{\Huge{j}} \raisebox{.1cm}{\phantom{\Huge{l}}}
Ref.~\cite{Colangelo1}& 0.21(7) & \parbox{5.3cm}{\footnotesize \sl Light cone SR (calcul. of {$g_{B^*B^*\pi}$}).} \\
\vspace*{-4mm}
&& \\ 
\hline \hline
\multicolumn{3}{|c|}{\sf Quark Models}   \\
\hline
\phantom{\Huge{l}} \raisebox{.05cm}{\phantom{\Huge{j}}}
Ref.~\cite{Yan}& 1.0 & \parbox{5.3cm}{\footnotesize \sl Non-relativistic QM.} \\
\phantom{\Huge{l}} \raisebox{.1cm}{\phantom{\Huge{j}}}
Ref.~\cite{Xu}& 0.60(5) & \parbox{5.3cm}{\footnotesize \sl Light front QM-{\sc I}. } \\
\phantom{\Huge{l}} \raisebox{.1cm}{\phantom{\Huge{j}}}
Ref.~\cite{Jaus}& 0.57 & \parbox{5.3cm}{\footnotesize \sl Light front QM-{\sc II}. } \\
\phantom{\Huge{l}} \raisebox{.1cm}{\phantom{\Huge{j}}}
Ref.~\cite{Deandrea}& 0.46(4) & \parbox{5.3cm}{\footnotesize \sl Model which includes HQS \\ and approximate $\chi S$.} \\
\phantom{\Huge{l}} \raisebox{.1cm}{\phantom{\Huge{j}}}
Ref.~\cite{Colangelo2}& 1/3 & \parbox{5.3cm}{\footnotesize \sl Partially relativistic model.} \\
\vspace*{-4mm}
&& \\ 
\hline
\phantom{\Huge{l}} \raisebox{.15cm}{\phantom{\Huge{j}}}
This work & \, {\sf 0.6(1)} & \parbox{5.3cm}{\footnotesize \sl  With the preferred set of parameters: $\widehat g = 0.61$ (see the text).} \\
\vspace*{-4mm}
&& \\ 
\hline \hline
\multicolumn{3}{|c|}{\sf ``Experiment''}   \\ 
\hline
\phantom{\Huge{l}} \raisebox{.01cm}{\phantom{\Huge{j}}}
Ref.~\cite{Barlag}&  $<\ 0.76$ & \parbox{5.2cm}{\footnotesize \sl Experimental upper \\ bound (see footnote~1).} \\
\phantom{\Huge{l}} \raisebox{.2cm}{\phantom{\Huge{lj}}}
\hspace*{-3mm} Ref.~\cite{Stewart}&  \quad \quad \parbox{2.0cm}{0.27$^{\ +4\ +5}_{\ -2\ -2}$\\ 0.76$^{\ +3\ +2}_{\ -3\ -1}$}$\Biggr\}_{\raisebox{-.15cm}{\phantom{\Huge{j}}}}$ & \parbox{5.2cm}{\footnotesize \sl From experimental data and \\ accounting for the HQS and $\chi$S.} \\
  \hline
\end{tabular}
\end{center}
\caption[]{\sl Results for $\widehat{g}$ as obtained from the lattice QCD, various QCD sum rules and several quark models. In Ref.~\cite{Stewart}, the lagrangian approach (reviewed in~\cite{Casalbuoni}) and the experimental results for ${\cal{B}}(D^{*+}\to D^+ \gamma)$ and ${\cal{B}}(D^{*+}\to D^0 \pi^+)$~\cite{PDG}, allowed the extraction of $\widehat{g}$, but two solutions were found. Along the same lines, but using the experimental ${\cal{B}}(D^0\to K^- \ell \nu_\ell)$ and $f_{D_s}$, two distinct solutions for $\widehat g$ were obtained also in Ref.~\cite{Fajfer}.}  \label{tab0}
\vspace*{-1cm}
\end{table} 
}

From the physical point of view, a question arises when one tries to reconcile the small values for $\widehat g$ with the Adler--Weisberger sum rule~(AWSR). The experience with the coupling to $\pi N$, taught us that lowest level resonances give the essential contribution to the saturation of the sum rule. This certainly would not be the case for $\pi B$ scattering if the value of $\widehat g$ were very small. Another general problem for a small $\widehat g$
is the additivity property which is expected to relate the axial couplings 
of light quarks inside the various hadrons (see below for more discussion). Based on the study of the decay of charmed baryons~\cite{Pirjol2}, where the coupling is known experimentally, it can be inferred that, in general, there is {\bf no} such a strong suppression (of $\widehat g$).

We want to rediscuss the problem within the quark model approach, to see whether one can get more stable predictions, restore the saturation of AWSR as discussed above, and retrieve the additivity. This is achieved by working in the Dirac equation framework\footnote{While this paper was in writing, a similar strategy to calculate the pionic couplings was also considered in Ref.~{Roberts}. However the subtleties of our models are quite different.}. After presenting the basic definitions and motivations in Sec.~\ref{one}, we discuss the potential and sketch the calculation in Sec.~\ref{two}. Discussion of the results are presented in Sec.~\ref{three}. We end with a short conclusion in Sec.~\ref{Conclusion}. 


\section{Dirac equation, pion emission and Adler--Weisberger sum rule}
\label{one}

A nice feature of the model based on Dirac equation in a central potential is that it treats relativistically the quark internal velocities, with a certain amount of precision. In particular, it includes the Z graphs and appreciably modifies the estimate of the small components as compared to free spinors. The latter is crucial for the estimate of quantities such as axial couplings. It presents exact additivity for the current matrix elements. This model was initiated by Bogolioubov~\cite{Bogolioubov}, who used it to calculate $g_A$ - nucleon axial coupling (related to the $\pi N$ scattering). In the soft pion limit, one encounters the famous AWSR:
\bea
\phantom{\Huge{l}} \raisebox{.3cm}{\phantom{\Huge{j}}}
g_A^2 \ -\  {f_\pi^2 \over \pi} \ \int {ds \over s - m_N^2} \left[\sigma(\pi^+p) - \sigma(\pi^-p)\right] \ =\ 1\ .
\eea

Bogolioubov obtained the result: $g_A = 5/3 \ (1-2\delta)$, where a positive $\delta$ is a measure of the small components of Dirac spinors. With large
quark velocities found in hadrons, the non-relativistic result $g_A = 5/3$ is significantly reduced, and a good agreement with experiment ($1.267(4)$~\cite{PDG}) was reached.
In the saturation procedure with the nucleon in the game, one should pay attention to quark's isospin and then sum over all quarks. In the case of meson
with a very massive flavor, the situation is even simpler since we deal with only one light quark~(heavy quark decouple due to the heavy quark symmetry). Thus, the situation is very favorable for the use of the Dirac equation, which precisely corresponds to a particle in a static center of force. This is not only true for the 
description of the spectrum, but also for a quantum emission from the light quark
(radiative transitions, pion emission/absorption)\footnote{Thus for the processes when the recoil of the heavy quark can be neglected.}.\\

The exact validity of the AWSR in this (Dirac) formalism has already been demonstrated in Ref.~\cite{Gavela}. We now recall some basic elements.
Consider the axial current
\bea
A_\mu^i (x)\ =\ \bar q(x) \gamma_\mu \gamma_5 {\tau^i \over 2} q(x)\ ,
\eea
and its corresponding charge
\bea
Q_A^i (n)\  =\  \int d^4 x \ \delta(n\cdot x) \ n^\mu A_\mu^i(x)\ ,
\eea
where $\tau^i$ are ordinary Pauli matrices, whereas $n=(n^0,\vec n)$ is a timelike unit vector. In the null-plane limit ($\vert\vec n\vert/n^0 \to 1$), the matrix element between one-quark states, {${\rm X}^i_{\alpha \beta}(n) \equiv \langle \alpha \ \vert Q_A^i(n) \vert \ \beta \rangle$}, describes the pionic transition ($m_{\pi}=0$):
\bea
\label{lab10}
{\rm X}^i_{\alpha \beta}(n) \ = \ \int d^4 x \ \delta(n\cdot x) \ \bar \Psi_{\alpha}(x) (n\cdot \gamma) \gamma_5 {\tau^i\over 2} \Psi_{\beta}(x)\ ,
\eea
\noindent
where $\Psi_i (x)$ represent a complete set of solutions of the Dirac equation, which we take to be the eigenstates ($\Psi_i (x) = \Psi_i(\vec r) e^{-i E_i t}$), whereas $n$ is chosen to be collinear with the pion four-momentum
\bea
\label{glavna}
\phantom{\Huge{l}} \raisebox{.5cm}{\phantom{\Huge{j}}}
&&{\rm X}^i_{\alpha \beta}(n) =\int d \vec r \ \bar \Psi_{\alpha}(\vec r) \left(\gamma_0 - {\vec n\over n^0}\vec \gamma \right) \gamma_5  {\tau^i\over 2} \Psi_{\beta}(\vec r)\  e^{-i (E_{\beta} - E_{\alpha}) \vec n \cdot \vec r / n^0} \\
&&\hfill \nonumber
\eea
The last step is to use $\left[ Q^+_A(n),Q^-_A(n)\right] = 2 I_3(n)$, demonstrated to be valid in the model~\cite{Gavela}, in the null-plane limit ($\vert \vec n\vert / n^0 \to 1$). $I_3(n)$ is the corresponding vector charge. One finally arrives at
\bea
\sum_\delta \ {\rm X}_{\alpha \delta}\ {\rm X}_{\delta \beta}\ =\ \delta_{\alpha \beta}\; ,
\eea
\noindent
{\it i.e.} the Weinberg-like, matrix form of the AWSR~\cite{Weinberg}. By specifying
the null-vector $n$ as $n^2=q^2\to 0$, we have rewritten~(\ref{glavna}) as: 
\bea
{\rm X}_{\alpha \beta}=  \int d\vec r \ \Psi^{\dagger}_\alpha(\vec r)(\gamma_5 - {\vec q\over q^0}\vec \Sigma)\Psi_{\beta}(\vec r)\  e^{-i\vec q~\vec r} \label{glavna2} \ ,
\eea
where $\vec \Sigma = {\bf 1}\otimes \vec \sigma$, as usual, and with $\tau$ matrices dropped out thanks to the additivity feature of the Dirac model. 
As we already mentioned, a heavy-light meson is (in the heavy quark limit) the one light quark system, and we can set $\vert \alpha \rangle = \vert \beta \rangle = \vert B\rangle$, and write
\bea
\sum_\delta \vert {\rm X}_{B\delta}\vert^2 = 1 \ .
\label{4} 
\eea
This is the form of the AWSR that we are interested in. ${\rm X}$'s will be directly related to the pionic decay widths through the Weinberg's formula ($m_{\pi}=0$):
\label{1} 
\bea 
\Gamma({\cal{I}} \, \to \, {\cal{F}}\  \pi)\ =\  {1\over  {\,\, 2 \pi\ f_\pi^2\,\, }} \, \frac{\vert \ \vec q\ \vert^3}{\,\, 2 j_{\cal I}  \ +\ 1\,\, } \, \sum \vert {\rm X}_{{\cal{I}} \to {\cal{F}}} \vert^2 ,
\label{2} 
\eea 
with sums over initial and final polarisations and over final charge states, and an additional factor $1/2$ for neutral pion ($f_{\pi}=131~{\rm MeV}$). This will be the main formula for our investigation.

We must emphasize at this stage, that a basic assumption we make is
that the QM current is identical to the bare one
(in terms of quark operators). Namely, the QCD radiative corrections in principle introduce a $(g_A)_q$ factor for quarks~\cite{Livre}, representing the phenomenological renormalization for constituent quarks.
We do not introduce such a factor for two reasons. {\underline First}, we want to
remain within the point of view of saturating the AWSR by states composed
of valence quarks only. It is rather obvious that the sum~(\ref{4}) 
will no more equate $1$,
if the axial quark current is multiplied by $(g_A)_q \not =1$. On the other hand, if one includes radiative corrections, then one should also introduce the states with gluons, to keep the consistency with AWSR. 
{\underline Second}, such a factor $(g_A)_q$ notably different from $1$, does not seem to be required to explain the (similar) charmed baryon decays, or the nucleon  $(g_A)_N$ (see Sec.~\ref{discussion2}).  At least, it could not be very small. Nevertheless, a reduction factor, not too different from $1$, is by no means excluded.

In phenomenological Lagrangians~(see \cite{Casalbuoni,AFalk} and references therein), the pionic transitions are basically described in terms of invariant coupling constants. This has several advantages: the phenomenological couplings incorporate explicitly Lorentz invariance and the consequences of the $J^P$
properties for the decay; they may also include the consequences of the $HQS$ and chiral symmetry~($\chi S$). From the $HQS$, we know that the wave function of a heavy meson is independent of the heavy flavor and its spin. Consequently, the total angular momentum of the light degrees of freedom ( {\it i.e.} of a light quark)  $j_\ell$, is a good quantum number and heavy mesons come in degenerate doublets with spin $J = j_\ell \pm \frac{1}{2}$. The doublets of interest here, are:  
\bea
\left[\ B(_{\frac{1}{2}}0^-)\ ,\ B^*(_{\frac{1}{2}}1^-)\ \right];\quad 
\left[\ B_0^*(_{\frac{1}{2}}0^+)\,\ B_1'(_{\frac{1}{2}}1^+)\ \right];\quad \left[\ B_1(_{\frac{3}{2}}1^+)\ ,\ B_2^*(_{\frac{3}{2}}2^+)
\right]\, ,
\eea
with the notation, $_{j_\ell} J^P$. From the parity conservation, one can easily see that: {\sf i)} the transition from one meson to the other within a doublet, occurs with the emission (absorption) of the P-wave pion; {\sf ii)} between $B$ and $B_0^*$, as well as between $B^*$ and $B_1(B_1')$, the transition occurs with the S-wave pion; {\sf iii)} between two mesons of next to neighbouring doublets, with the D-wave pion (when allowed), and so on. The covariance properties of these transitions are parametrized by the invariant couplings, $g_{B_i B \pi}$, with B denoting here generically heavy-light meson, and $B_i$ stands for $B^*,\, B_0^*,\, B_2^*$ and (in general) for their radial excitations too. We will henceforth leave aside the $1^+$ states to avoid the complications of mixing, which anyway cannot be described by the Dirac equation. 
\bea
\label{coupl}
\lan B\  \pi \ \vert \ B^*\ran &=& ( q_\mu \ \epsilon^\mu )\, g_{B^* B \pi}\cr
&& \cr
\lan B\  \pi\  \vert \ B^*_0 \ran &=& g_{B^*_0 B \pi} \cr
&& \cr
\lan B\  \pi\  \vert\  B^*_2 \ran &=& \sqrt{\ {2\over 3}\ }\ (\epsilon_{\mu \nu}q^\mu q^\nu)\  g_{B^*_2 B \pi}
\eea
$q_\mu$ is the pion four momentum, whereas $\epsilon_\mu (\epsilon_{\mu \nu})$ is the polarization vector (tensor) of $B^* (B_2^*)$. In this definition $g_{B^* B \pi}\equiv g_{ B^{* 0} B^- \pi^+}$, is related to the other charge combinations by isospin:
\bea
g_{B^* B \pi}\ =\ g_{B^{* 0} B^- \pi^+}\ =\ -g_{B^{*-}\bar B^0 \pi^-}\ =\ \sqrt{2} g_{B^{*0}B^0\pi^0}\ =\ -\sqrt{2}g_{B^{*+}B^+\pi^0}\ .
\eea
$g_{B^*_0 B \pi}$ and $g_{B^*_2 B \pi}$ have the mass dimension $[m]$ and $[m^{-1}]$, respectively. Obviously, the above definitions apply in the same way to $D$ and $B$ mesons. The corresponding decay widths are: 
\bea
\Gamma ( B^{*0} \to B^+ \pi^- ) &=& \frac{1}{\ 24 \pi\ }\ g_{B^* B \pi}^2 \ \frac{\vert  \ \vec q \ \vert^3}{\ m_{B^*}^2\ } \,,\cr
&& \cr
\Gamma ( B^{*0}_0 \to B^+ \pi^- ) &=& \frac{1}{\ 8 \pi\ }\  g_{B^*_0 B \pi}^2\  \frac{\vert\  \vec q\ \vert}{\ m_{B^*_0}^2\ } \,,\cr
&& \cr
\Gamma ( B^{*0}_2 \to B^+ \pi^- ) &=& \frac{1}{\ 40 \pi\ }\  g_{B^*_2 B \pi}^2\  \frac{\vert\  \vec q\ \vert^5}{\ m_{B^*_2}^2\ }\,, 
\label{3}
\eea
By writing this generically, we have~{\footnote{Remark again that unlike for $\Gamma(D^* \to D \pi)$, the case, $B^* \to B \pi$, is kinematically forbidden.}}
\bea
\Gamma ( B^{*0}_i \to B^+ \pi^- ) &=& \frac{1}{\ 8 \pi\  (2 j_{B^*_i} + 1)\ }\  g_{B^*_i B \pi}^2 \ \frac{\ \vert\  \vec q\ \vert^{2 l + 1}}{m_{B^*_i}^2\ }\,
\eea
with $\ell = j_{B^*_i}$ being the spin of the decaying meson. If one includes the constraints of $HQS$ and $\chi S$ in the phenomenological lagrangians~\cite{Casalbuoni,AFalk}, the pionic transitions are described (at lowest order) by the new, dimensionless couplings: $(g=\widehat g)$, $h$, $h'$, which are the crucial parameters of the approach and are independent of the heavy mass. In fact, the X's which we defined above~(\ref{glavna2}), are essentially these couplings\footnote{Indeed, the above X's are defined for the Dirac equation, and will be valid in the heavy mass limit ($m_Q \to \infty$) of the theory.}. To obtain the exact relations with our X's, we compare the corresponding widths in the heavy quark limit. In terms of the new couplings, the expressions~(\ref{3}) read \footnote{For the $B^* \to B$ transition, our formula derived exactly from  the Lagrangian differs by the factor $m_B/m_{B^*}$
from the usual expression.}
\bea
\phantom{\large{l}} \raisebox{.5cm}{\phantom{\large{j}}}
\Gamma (B^{*0} \to B^+ \pi^-) &=& \frac{{\widehat g}^2}{\ 6 \pi f_\pi^2\ }\ \frac{m_B} {m_{B^*}}\vert \ \vec q\ \vert^3 \,, \cr
&& \cr
\phantom{\large{l}} \raisebox{.5cm}{\phantom{\large{j}}}
\Gamma(B^{*0}_0 \to B^+ \pi^-) &=& \frac{{h}^2}{\ 8 \pi f_\pi^2\ }\  {m_B\over m_{B_0^*}^3}\ (m_{B^*_0}^2 - m_B^2)^2\  \vert \ \vec q\ \vert \,, \cr
&& \cr
\phantom{\large{l}} \raisebox{.5cm}{\phantom{\large{j}}}
\Gamma (B^{*0}_2 \to B^+ \pi^-) &=& \frac{{h'}^2}{\ 15 \pi f_\pi^2 \Lambda_\chi^2\ }\ \frac{m_B}{\  m_{B^*_2}\  }\  \vert \ \vec q\ \vert^5 \, . 
\label{couplingwidths2}
\eea

On the other hand, we know that the soft pion emission expressed in terms of $X$'s is~(\ref{2})
\bea
\Gamma(B^{*0}_i \to B^+ \pi^-) = \frac{1}{\ 2 \pi f_\pi^2\ }\  \frac{\vert \ \vec q\ \vert^3}{\ 2 j_{B^*_i} + 1\ }\  \vert \ {\rm X}_i\  \vert^2 \,,
\eea
with ${\rm X}_i$ calculated for the $u$ quark and zero polarisations, so that we can identify (up to phases):
\bea
g_{B^* B \pi}\ =\ {2\ \sqrt{\phantom{\large{l}} m_B \ m_{B^*}\ } \over f_\pi}\ g &;&\hspace*{1.5cm} g=\widehat g = {\rm X}_{B^*}\,,\cr
&& \cr
\phantom{\large{l}} \raisebox{.5cm}{\phantom{\large{j}}}
g_{B_0^* B \pi}\ =  \ \sqrt{\frac{\textstyle{\large m_B }}{ \textstyle{\large m_{B^*_0}}}}\ (m_{B^*_0}^2 - m_B^2)\ {\ h\ \over f_\pi}  &;&\hspace*{1.5cm} h = {\rm X}_{B_0^*}\,,\cr
&& \hfill\cr
\phantom{\large{l}} \raisebox{.7cm}{\phantom{\large{j}}}
g_{B_2^* B \pi}=\sqrt{\ 8\ \over 3}\ {\sqrt{\phantom{\large{l}} m_{B_2^*}\ m_B\ } \over f_{\pi}}\ {h'\over \Lambda_\chi}&;&\hspace*{1.5cm}{h'\over \Lambda_\chi} = \sqrt{\ 3\ \over 2}\ {{\rm X}_{B_2^*} \over \Delta E} \,.
\label{gX}
\eea
The kinematical factors in the first column are
 kept to their physical values, to (somewhat) account for finite mass effects, although the higher~(in $1/m_Q$) couplings are being neglected. On the other hand, the second column is obtained consistently in the heavy quark limit, since the couplings $g,\ h,\ h'$ are the only dominant in this limit. X's are to be obtained from the Dirac equation (in the $m_Q\to \infty$ limit), where we will also assume $m_{\pi}=0$, i.e. $\omega_{\pi}=|\ \vec q\ |=\Delta E$ ($\Delta E$ is the difference between the final and initial Dirac energy level). 
$\Lambda_\chi\simeq 1\,{\rm GeV}$, is the chiral symmetry breaking scale.
Admittedly, the choice we make is somewhat arbitrary, although conformal to our experience with quark models. Another reasonable procedure which can be used, consists in fixing the matrix element $\omega_{\pi}~{\rm X}/f_{\pi}$, to its
limiting value, $\Delta E~{\rm X}/f_\pi$, and to take only the phase space factor at its physical value. The resulting widths are somewhat different which indicates the uncertainty in the results.

 From the above relations, we see that (in the $m_Q\to \infty$ limit) the AWSR~(\ref{4}) can be rewritten in the following form:
\bea
(g=\widehat{g})^2 \underbrace{+ {h}^2 + {2\over 3} (\Delta E)^2 {h'^2 \over \Lambda_\chi^2} + \dots}_{\parbox[b]{4.2cm}{\vspace*{-3mm} \begin{center} \small \sl orbital excitations\end{center}}}
 \ + \, \parbox{2.2cm}{\begin{center} \sl radial\\ excitations\end{center}}\ + \ \parbox{2.6cm}{\begin{center} \sl multiparticle\\  states\end{center}} \ =\  1\, .\quad
\label{10}
\eea

\vspace{.8cm}

\section{Calculations with a specific class of Dirac equation}
\label{two}
\subsection{Potential and solution of Dirac equation}

To make use of (\ref{glavna}), we have to find eigenstates of the Dirac equation: 
\bea
\biggl(\ \vec \alpha \cdot \vec p \ +\ m \beta \ +\ V(\vec r)\ -\ E \biggr)\, \Psi (\vec r) = 0
\label{DEQ}
\eea
and this is the part where we have to fix the model. Models based on Dirac equation differ in the form of the potential $V(\vec r)$. In particular, the differences come from the Lorentz structure of the confining part of the potential. Here, we will take the linear confining potential as a {\sl Lorentz scalar}:
\bea
V(r)\ =\ - {\ \kappa\  \over r}\ +\ \beta \ \left( a r +c \right)\,.
\label{7}
\eea
\noindent The Coulomb constant $\kappa$, the string tension $a$, and the constant $c$ (which is included in the scalar-confining part~\footnote{If a constant $c$ were included in the vector part, the resulting effect would be an overall shift of the spectrum.}), are the parameters of the potential. Note that
only the combination $\left( m+c \right)$ enters the equation~(\ref{DEQ}). Therefore, there will be only three parameters
in the model: $a$, $\kappa$ and $\left( m+c \right)$. It must be
emphasized that neither $m$, nor $\left( m+c \right)$ corresponds
to the constituent quark mass, so that neither
should be constrained to remain around 300 MeV; $\left( m+c \right)$
may even be negative ({\sl see below}). In Ref.~\cite{Olsson}, this model was discussed with various forms of the potential. On theoretical grounds (see~\cite{Mur,Vairo}), the  Lorentz structure of the confining potential is a delicate issue. There is no reason to take it as a vector, since it cannot correspond to the exchange 
of one single gluon, even if dressed by very large self-energy effects. 
Indeed, general considerations for the heavy quark-antiquark potential 
lead to a spin-orbit force with a sign opposite to the one predicted by 
a vector exchange~\cite{Gromes}. Yet, this does not allow to say that the confining potential is the Lorentz scalar. The scalar is only the simplest choice which is in agreement with the quarkonium phenomenology, and in particular with the spin structure (which is very sensitive to the Lorentz assignment of the potential). Therefore, it is reasonable to calculate our X's adopting such a potential~(\ref{7}). 
If the constant $c$ is included in the scalar confining part (and there is {\em a priori} no reason to discard it), only the combination $( m + c )$ (and not only $m$) enters the game. In other words, we cannot distinguish
between the `quark mass' and the constant in the potential. This is not
worrisome, because $m$ does not play the role of the constituent mass, in the framework of the Dirac model (see below). \par

For the solution of the Dirac equation, we will heavily rely on the work of the Wisconsin group on the $D,~D_s,~B,~B_s$ (spin averaged) states~\cite{Olsson}, where they also discussed the way how to numerically treat the Dirac equation with the potential. We differ however on the questions such as the string tension, the quark mass and the constant in the potential (which we allow to be non-zero).

The general solution of the spherically symmetric problem at hand, can be written as:
\bea
\hspace*{-3mm}\phantom{\Huge{l}} \raisebox{6mm}{\phantom{\Huge{j}}}
\Psi_{jm}^{k}({\vec r})&=&
			\left(\begin{array}{c}
\hspace*{-3mm}\phantom{\Huge{l}} \raisebox{-.2cm}{\phantom{\Huge{j}}}
f_{j}^{k}(r)\ {\cal Y}_{jm}^{k}({ \widehat{r}})\, \\
\hspace*{-3mm}\phantom{\Huge{l}} \raisebox{.0cm}{\phantom{\Huge{j}}}
ig_{j}^{k}(r)\ {\cal Y}_{jm}^{-k}({ \widehat{r}})\, 
\end{array}\right) \ =\ 
\left(\begin{array}{cc}
\hspace*{-3mm}\phantom{\Huge{l}} \raisebox{-.2cm}{\phantom{\Huge{j}}}
	\ {\cal Y}_{jm}^{k}({ \widehat{r}})\  & 0 \\
\hspace*{-3mm}\phantom{\Huge{l}} \raisebox{.0cm}{\phantom{\Huge{j}}}
	\ 0\  		  & \ i{\cal Y}_{jm}^{-k}({ \widehat{r}})\ \end{array}\right)\, \left(\begin{array}{c}
\hspace*{-3mm}\phantom{\Huge{l}} \raisebox{-.2cm}{\phantom{\Huge{j}}}
  		f_{j}^{k}(r) \\
\hspace*{-3mm}\phantom{\Huge{l}} \raisebox{.0cm}{\phantom{\Huge{j}}}
		g_{j}^{k}(r) \end{array}\right) \ .
\eea
\vspace*{.1cm}\\
\noindent
${\cal Y}_{jm}^{-k}({ \widehat{r}})$ denote usual spherical harmonics, where $k$ stands for the Dirac quantum number ($k=l$ for $l=j+\frac{1}{2}$, and $k=-(l+1)$ for $l=j-\frac{1}{2}$). After inserting this into the Dirac equation, one arrives at the system of differential equations,
\bea
\hspace*{-3mm}\phantom{\Huge{l}} \raisebox{.6cm}{\phantom{\Huge{j}}}
\left(\begin{array}{cc}
\hspace*{-3mm}\phantom{\Huge{l}} \raisebox{-.2cm}{\phantom{\Huge{j}}}
	m  + a r + c - { \textstyle{\large \kappa} \over \textstyle{\large r}} - E & {\textstyle{\large k}\over \textstyle{\large r}} - \left( \frac{\textstyle{\large d}}{\textstyle{\large dr}} + {\textstyle{\large 1}\over \textstyle{\large r}}\right) \\
\hspace*{-3mm}\phantom{\Huge{l}} \raisebox{.1cm}{\phantom{\Huge{j}}}
	{\textstyle{\large k}\over \textstyle{\large r}} + \left( \frac{\textstyle{\large d}}{\textstyle{\large dr}} + {\textstyle{\large 1}\over \textstyle{\large r} } \right) & \quad - \left( m + a r + c \right) -  {\textstyle{\large \kappa} \over \textstyle{\large r}} - E  \end{array} \right) \, \left(\begin{array}{c}
\hspace*{-3mm}\phantom{\Huge{l}} \raisebox{-.2cm}{\phantom{\Huge{j}}}
  		f_{j}^{k}(r) \\
\hspace*{-3mm}\phantom{\Huge{l}} \raisebox{.1cm}{\phantom{\Huge{j}}}
		g_{j}^{k}(r) \end{array}\right) = 0\, .
\label{12}
\eea
\vspace*{.1cm}\\
\noindent
As it was proposed in \cite{Mur}, we look for the solutions of the radial part
\bea
f_{j}^{k}(r)\simeq\sum_{i=1}^{N}\ c_{i}^{(f)}\ \phi_{il(k)}\ ,
\hspace*{2cm}
g_{j}^{k}(r)\simeq\sum_{i=1}^{N}\ c_{i}^{(g)}\ \phi_{il(k)}\ .
\label{8}
\eea
in the pseudocoulombic basis of functions:
\bea
\hspace*{-3mm}\phantom{\Huge{l}} \raisebox{.5cm}{\phantom{\Huge{j}}}
\phi_{il}(r)\ =\ \sqrt{\frac{8 (i-1)!}{(i+2l+1)!}}\ \beta^{\frac{3}{2}}\ (2\beta r)^{l}\ e^{-\beta r}\ 
L_{i-1}^{2l+2}(2\beta r)\ . 
\label{9}
\eea
\vspace*{.1cm}\\
\noindent
For a certain $N$, one diagonalizes the hamiltonian and gets $N$ positive and $N$ negative energy solutions. The eigenvalues $E_i$, should not depend on the variation parameter $\beta$, which is actually the case if $N$ is large enough (in our case it turns out to be for $N\geq 15$). In order to fix the parameters of the potential,
one has to compare the predicted spectrum with the experiment, which is not a trivial task. The Dirac equation has the properties of heavy quark mass limit, {\it i.e.} the mass of a heavy-light meson is $m_H = m_Q + E_\ell+{\cal O} (1/m_Q)$. If we neglect the terms ${\cal O} (1/m_Q)$,
$m_Q$ can be adjusted trivially for each quark species to fit the ground state, or more precisely, the spin-averaged ($0^{-},1^{-}$) ground state. However, this approximation appears to be too crude, as we show below.\par
Let us reconsider the problem of comparing the 
experimental spectrum with the predictions
of the Dirac equation.  Note that in the latter only the spin-orbit force (relative to the light quark) survives, whereas the spin-spin force is not present\footnote{This implies that {\it e.g.}, the $L=1$ states with same $j_\ell$ are degenerate.}. What is really predictable and comparable with experiment is: $\Delta E_1^\infty$, the mass difference between the spin average of the $(L=1,j_\ell=3/2)$ states and the spin-averaged ($0^{-},1^{-}$) ground state (extrapolated to the heavy quark limit), and $\Delta E_{s.o.}$, the spin-orbit splitting between $(L=1, j_\ell=3/2)$ and $(L=1,j_\ell=1/2)$ states (in principle also to be extrapolated).\par

$(\Delta E_1^\infty)^{\rm (exp)}$ is known, modulo spin averages and an extrapolation in
$1/m_Q$. By using $D$ and $B$ states~\cite{PDG} ({\it i.e.} $\Delta E_1^c$ and $\Delta E_1^b$) and extrapolating as
\bea
\Delta E_1^Q \ =\ \Delta E_{1}^{\infty}\ +\ {{\cal A} \over \ m_Q\ }\, ,
\label{6}
\eea
one finds that the coefficient ${\cal A}$ is sizable ${\cal A}\simeq 0.16~{\rm GeV}^2$. That means that for $D$ mesons, the corrections are as large as $\sim \ 25\%$.\par
As for the $L=1,j_\ell=1/2$ states, they are yet to be detecteded, although an interesting constraint on the spin-orbit splitting can be deduced from existing data. 
Namely, the observed (broad) $B^{**}$ bump~\footnote{OPAL Coll.~\cite{OPAL}, and DELPHI Coll.~\cite{DELPHI} reported $\Gamma(B^{**})= 116(24)~{\rm MeV}$ and $\Gamma(B^{**})= 145(28)~{\rm MeV}$, respectively.} cannot be {\sl a priori} ascribed to the (narrow) $j_\ell = 3/2 $ states only. That means that the
$L=1,j_\ell=1/2$ states are also present in $B^{**}$, and their widths should be around $100~{\rm MeV}$ (which may be the explanation why they were not isolated).
This fact gives an important,
additional constraint on the spectrum: the splitting between $j_\ell=3/2$ and $j_\ell=1/2$ states must be small, {\it i.e.} such that only one bump stays `visible'. \par
Thus, the two firm experimental facts which we can compare with the predictions of the Dirac equation (and thus fix the parameters of our model) are: the value of $\Delta E_1^{\infty}$, and
the ({\sl small}) spin-orbit splitting, $\Delta E_{s.o.}$. \\
\noindent
 The set of parameters which reproduced the spectrum in \cite{Olsson} was:
\bea
m \ =\ 300~{\rm MeV}\ ,\quad \, a\ =\ 0.308~{\rm GeV}^2\ ,\quad \kappa \ =\  0.579\ , 
\eea
with constant $c$ set to nought. The manifest problem of such a model is the large value of the string tension, which is normally expected to be, $0.15 \lesssim a \lesssim 0.20~{\rm GeV}^2$. After inspection, we realized that the problem appeared because they fit the spectrum of the $D$ states ($\Delta E_1^c = 0.47~{\rm MeV}$), while the Dirac equation is valid in the infinite mass limit. From the measured $B^{**}$ (L=1) states, we can
infer $\Delta E_1^b \simeq 0.38~{\rm MeV}$, which means that it should be even smaller when $m_Q \to \infty$. Linear extrapolation in Eq.~(\ref{6}) amounts to $\Delta E_1^\infty \approx 0.35~MeV$, close to the existing estimate~\cite{Mehen}. 
In order to fit this ({\sl smaller}) value of $\Delta E_1^\infty$, a much lower value of the string tension $a$
is needed than that found by authors of Ref.~\cite{Olsson}. In other words, the string tension is very sensitive to the value of $\Delta E_1^\infty$. 
Since $\Delta E_1^\infty < \Delta E_1^c$, we can fit the spectrum with
standard values for $a$. Reasonable values of the other parameters are $-0.2\lesssim ( m + c) \lesssim 0.3$ and $\kappa \gtrsim 0.6$, the simultaneous choice of the three parameters must be correlated. Our preferred choice, which was used to get the results listed in Tabs.~(\ref{tab1},\ref{tab2},\ref{tab3}), is:
\bea
- ( m + c ) = 200~{\rm MeV}\ ,\quad \, a=0.20~{\rm GeV}^2\ ,\quad \kappa = 0.65\ . \label{19}
\eea

\begin{table}[hbt]
\begin{center}
\begin{tabular}{|c|c|c|} 
\cline{1-1} \cline{3-3}
\phantom{\Huge{l}} \raisebox{.2cm}{\phantom{\Huge{j}}}
\hspace*{-5mm}{\underline{\sf Parameters of the potential}~(Eq.~\ref{7})}& \hspace*{1.5cm}&{\hspace*{1.1cm} \underline{\sf  Dirac spectrum} \hspace*{1.1cm} }\\
\phantom{\Huge{l}} \raisebox{.1cm}{\phantom{\Huge{j}}}
$ a \ =\ 0.20~{\rm GeV}^2$& & $\Delta E_1^\infty = 341~{\rm MeV}$\\
\phantom{\Huge{l}} \raisebox{.1cm}{\phantom{\Huge{j}}}
$\kappa\ =\ 0.65$ & & $\Delta E_{s.o.}= 1.3~{\rm MeV}$\\
\phantom{\Huge{l}} \raisebox{.1cm}{\phantom{\Huge{j}}}
$-( m + c)\ =\ 200~{\rm MeV}$& & $\Delta E'= 494~{\rm MeV}$\\

\cline{1-1} \cline{3-3}
\end{tabular}
\end{center}
\caption[]{\sl Our preferred set of parameters used for the spectrum for which we tabulate: 
$\Delta E_1^\infty = E_{Dirac}(L=1, k=-2) - E_{Dirac}(L=0)$ ; the small energy splitting within $L=1$ states due to the spin-orbit forces in the $m_Q=\infty$ limit, is $\Delta E_{s.o.} = E_{Dirac}(k=-2) - E_{Dirac}(k=+1)$, and $\Delta E'$ is the distance between the ground state and the first radial excitation in the same limit, also given by the Dirac equation.}
\label{tab1}
\end{table} 

\subsection{Expressions for pionic transitions}

Now we go back to our problem and calculate the pionic transitions from the obtained wave functions, solution of the above Dirac equation~(\ref{DEQ}). To end that, we take the pion emitted along the $z$-axis, and use the plane wave expansion in~(\ref{glavna2}), to end up with
\bea
\label{99}
\hspace*{-6mm}\phantom{\Huge{l}} \raisebox{.6cm}{\phantom{\Huge{j}}}
{\rm X}_{BB_n}\ = \ \sum_{l=0}^{\infty} (2 l + 1) (-i)^l \sqrt{\frac{4 \pi}{2 l + 1}} \int d\vec r\ \Psi_B^\dagger(\vec r)(\gamma^5 - \Sigma_3)\Psi_{B_n}(\vec r) j_l(q r) Y_{l}^{0}(\widehat r)\ .
\eea
\vspace*{.1cm}\\
\noindent
In terms of the eigenfunctions of the Dirac equation, the couplings (for transitions from excited to the ground state) read as follows: 
\bea
\hspace*{-3mm}\phantom{\Huge{l}} \raisebox{.5cm}{\phantom{\Huge{j}}}
{\rm X}_{BB^*}&=& \int_{0}^{\infty}\left\{ \left[ \vert f_{1/2}^{(-1)}\vert^2 - \frac{1}{3} \vert g_{1/2}^{(-1)}\vert^2   \right] j_0(qr) - \frac{4}{3}\  \vert g_{1/2}^{(-1)}\vert^2 j_2 (qr)\right\}\ r^2 d r\cr 
\hspace*{-3mm}\phantom{\Huge{l}} \raisebox{.7cm}{\phantom{\Huge{j}}}
{\rm X}_{BB^*_0}&=& \int_{0}^{\infty}\left\{ \left[ f_{1/2}^{(-1)*} g_{1/2}^{(1)} - g_{1/2}^{(1)*} f_{1/2}^{(-1)}\right] j_0(qr) +  \left[ f_{1/2}^{(-1)*} f_{1/2}^{(1)} + g_{1/2}^{(1)*} g_{1/2}^{(-1)}\right] j_1(qr)\right\} r^2 d r \cr
\hspace*{-3mm}\phantom{\Huge{l}} \raisebox{.7cm}{\phantom{\Huge{j}}}
{\rm X}_{BB^*_2}&=&  \sqrt{2}  \int_{0}^{\infty} \left\{ \left[ f_{1/2}^{(-1)*} f_{3/2}^{(-2)} - \frac{1}{5} g_{1/2}^{(-1)*} g_{3/2}^{(-2)}\right] j_1(qr) -  \left[ f_{1/2}^{(-1)*} g_{3/2}^{(-2)} - g_{1/2}^{(-1)*} f_{3/2}^{(-2)}\right] j_2(qr) \right.\cr
& &\hspace*{1cm}\left. - \frac{6}{5} \ g_{1/2}^{(-1)*} g_{3/2}^{(-2)} j_3(qr)\right\} r^2 d r\,.
\hspace*{-3mm}\phantom{\Huge{l}} \raisebox{.5cm}{\phantom{\Huge{j}}}
\eea
\noindent
For eigenvectors, the dependence on $r$ is implicit ($f^{(k)}_{j_\ell}\equiv f^{(k)}_{j_\ell}(r)$ and $g^{(k)}_{j_\ell}\equiv g^{(k)}_{j_\ell}(r)$), in the above formulae. In a given approximation, the same expressions apply for the radial excitations too, where the corresponding radial functions ($f$ and $g$) are eigenvectors of the corresponding eigenvalue, $E_{1/2}'=795~{\rm MeV}$, of the hamiltonian~(\ref{12}), in the basis~(\ref{8}) with $N=30$. Note again that we take $q=\vert \ \vec{q}\ \vert = \Delta E$; therefore, $\vert \ \vec q\ \vert =0$, for $D^* \to D\pi$ transition. The resulting values for the X's and the couplings~{$g, h, h'$}, are listed in Tab.~\ref{tab2}, whereas the corresponding decay widths are tabulated in Tab.~\ref{tab3}. \par
\begin{table}[hbt]
\begin{center}
\begin{tabular}{|c|} 
\hline \\
{\underline{\sf Ground state ($_{j_\ell}L = _{1/2}0$) elastic transition}} \\
\hfill \\
\hfill \\
{\small \sl Preferred parameters}\\ 
$a=0.20\ {\rm GeV}^2 ,\ \kappa=0.65,\ ( m + c) = - 0.2\ {\rm GeV}$\\
\hfill \\
\fbox{$X_{_{1/2}0 \to _{1/2}0}\ =\ \widehat{g}\ =\  0.607$}\\
\hfill \\
\hfill \\
{\small \sl Another set of parameters with satisfactory spectrum}\\ 
$a=0.15\ {\rm GeV}^2 ,\ \kappa=0.62,\ ( m + c) = 0.3\ {\rm GeV}$\\
\hfill \\
$X_{_{1/2}0 \to _{1/2}0}\ =\ \widehat{g}\ =\  0.728$\\
\hfill \\
\hline
\\
{\underline{\sf Transitions to ($L=1$) excited states}}\\
{\small \sl Preferred set of parameters}\\
\hfill \\
$|X_{_{1/2}0 \to _{1/2}1}|\ =\ |h|\ =\ 0.536 $\\
\hfill \\
$|X_{_{1/2}0 \to _{3/2}1}|\ =\ 0.234\ ,\ \to |h' / \Lambda_\chi| =\  0.841$\\
\hfill \\
\hline
\end{tabular}
\end{center}
\caption[]{\sl Transition matrix elements {\rm X}'s, and couplings of heavy mesons with a single pion. The basis of functions~(\ref{8},\ref{9}) is used with {\rm N=30}, and $(g_A)_q=1$, as discussed in the text.}\label{tab2}
\end{table} 

\begin{table}[t]
\begin{center}
\begin{tabular}{|c c c|} 
\hline 
& &\\
$g_{B^*B\pi}=49.1$&{\hspace*{15mm}}&$g_{D^*D\pi}=18$\\
& &\\
$\Gamma(B^* \to B \pi)={\rm NO}$&&$\Gamma(D^* \to D \pi)= 93.7\ {\rm keV} $\\
\hline
& &\\
$g_{B_0^*B\pi} \simeq 19.5\ {\rm GeV}$&&$g_{D_0^*D\pi}\simeq 8.2\ {\rm GeV}$ \\
& &\\
$\Gamma(B_0^* \to B \pi) \simeq 290\ {\rm MeV}$&&$\Gamma(D_0^* \to D \pi) \simeq 320\ {\rm MeV}$ \\
\hline
 & &\\
$g_{B_2^*B\pi}=58\ {\rm GeV}^{-1}$&&$g_{D_2^*D\pi}=22.5\ {\rm GeV}^{-1}$\\
& &\\
$\Gamma(B_2^* \to B^{(*)} \pi)=27\ {\rm MeV}$&&$\Gamma(D_2^* \to D^{(*)} \pi)=47\ {\rm MeV}$ \\
\hline 
\end{tabular}
\end{center}
\caption[]{\sl Decay widths, calculated in the way described in Sec.~\ref{one} ; we use the masses $m_{D_0^*}=2.4 \ {\rm GeV}$, $m_{D_2^*}=2.46 \ {\rm GeV}$, $m_{B^{**}}=5.732 \ {\rm GeV}$. Note that the widths are sensitive to the assumptions concerning excitation masses.}\label{tab3}

\end{table} 

\section{Discussion of the results}
\label{three}
\subsection{Discussion of the numerical estimate of $\widehat{g}$, comparison with QCD sum rules and experiment} \label{discussion1}
\hskip \parindent For $\vert \ \vec q \ \vert \ =\ 0$, the expression for {X$=\widehat{g}$} is particularly simple:
\bea
 & & \cr
 & &\widehat{g}\ =\ {\rm X}_{BB^*}
\ =\ \int_{0}^{\infty}\left\{ \, \vert f_{1/2}^{(-1)}\vert^2 \ -\  \frac{1}{3} \vert g_{1/2}^{(-1)}\vert^2  \, \right\}\  r^2 d r\, , 
\eea
with the normalisation condition,
\bea
 & & \cr
1&=& \int_{0}^{\infty}\left\{ \ \vert f_{1/2}^{(-1)}\vert^2 +\vert g_{1/2}^{(-1)}\vert^2   \ \right\}\ r^2 d r\, .
\eea
In the non-relativistic limit, the {\em small component}, $g_{1/2}^{(-1)}$ is zero, and therefore:
$\widehat{g}=1$. \\
This result would correspond to the famous $5/3$ for the nucleon.
As we already explained, with the small components different from zero, the value for $\widehat{g}$ is considerably reduced. In abstracto, it could be very small, but we obtain (with our preferred parameters),\\
\vspace*{-1cm}
\begin{center}
\framebox[3.5cm]{\rule{0cm}{0.6cm}{\raisebox{.4ex}[1.5ex][.75ex]{$\widehat{g} \simeq 0.61\,,$}}}
\end{center}
which is by far larger than the values advocated by QCD sum rules $\widehat{g} \sim 0.2 \div 0.3$. 
We stress that this moderate suppression in our model remains such, even when we vary the parameters in their reasonable ranges (see Sec.~\ref{two}). We obtain, $\widehat g \simeq 0.6\ \pm \ 0.1$. This value corresponds to what is needed to get $g_A/g_V < 5/3$, {\it i.e.} closer to experiment~\cite{Bogolioubov}. To better monitor the consistency, we also observe that if we adopt the free Dirac spinors and $m=0$, we get a result of~\cite{Colangelo2}, $\widehat{g}=1/3$, but the success of Bogolioubov's result for $g_A/g_V$ would be totally lost. 
In fact, in this argument, we invoke the general principle of additivity for the coupling of light quarks within different hadrons, which is exact in the 
framework of the Dirac equation : the light quarks are considered to be independent and bound to a common center of force. In the nucleon, it is possible that the equal light quark mass situation implies smaller internal velocities than in the $D$ mesons, which could invalidate the use of our Dirac model and the additivity. However, the additivity argument works well for light quarks in charmed baryons which are more analogous to $D$ mesons, and one finds indeed that the suppression with respect to the naive model seems experimentally moderate~\cite{Pirjol2}\footnote{We stress again that $g_A/g_V$ and $\widehat{g}$, should
be also suppressed by the QCD radiative corrections~\cite{Livre}, and therefore 
the small component correction must not be too large.}.\par

As for the sensitivity of our (relatively large) result on the choice of parameters, we emphasize that we have quite clear conclusion that $\widehat g \gtrsim 0.5$, regardless of the choice of parameters provided they are compatible with the spectrum. $\widehat g$ decreases with 
$( m + c )$, but the latter should be kept $\geq -0.2~{\rm GeV}$ to ensure that the spectrum is well reproduced. We repeat, in the framework of the quark model with Dirac equation, low values of $\widehat g $, like $0.2$ or $0.3$, are completely excluded.

On the other side, to keep $\widehat g $ compatible with the experimental upper bound (see Tab.~\ref{tab0}), we have to ensure that $( m + c ) \leq 300~{\rm MeV}$\footnote{With $( m + c ) \simeq 300~{\rm MeV}$, we have $\widehat{g} \simeq 0.73$, and our numbers are compatible with the bound, but quite close to it.}. 
With $( m + c )$ small and negative, the spectrum can be very well described, and we are safely below the experimental bound. This determined our preferred set~(\ref{19}). 
We must note that ${\cal O}(1/m_c)$ enhancement with respect to the $m_Q\to \infty$ prediction is expected. In~\cite{Burdman}, it is shown that
\bea
\left( {g_{B^*B\pi}\over m_Q}\right)\ \left(f_{B^*}\sqrt{m_Q}\right)\ =\  {\cal{C}}_{\infty}\ \left(1\ +\ {\cal{O}}\left({1\over m_Q^2}\right)\right)
\eea
where ${\cal{C}}_{\infty}$ is the $m_Q\to \infty$ limit of the {\it l.h.s.} Since $f_{B^*}\sqrt{m_Q}$, deviates from its limit by a {\em negative} ${\cal O}(1/m_Q)$ term, one sees that $g_{B^*B\pi}/m_Q$ deviates from its limit $\widehat{g}/f_{\pi}$ by a {\em positive} ${\cal{O}}(1/m_Q)$ term. This is not in contradiction with our way of extrapolating $g_{B^*B\pi}/m_Q$ (eq. \ref{gX}), since $\sqrt{m_B m_{B^*}}/m_Q$ may be expected to be larger than one.\\

\subsection{Interpretation of the results of the Dirac equation model in comparison with previous quark model calculations}

An intriguing question for quark modelists is why our findings are 
so different from the ones of Ref.~\cite{Colangelo2} ({\it i.e.} $\widehat{g}=1/3$). 
We already showed that this number can be obtained with a free Dirac spinor structure and
$m=0$, independently of the spatial wave function. This would correspond to 
\bea
\int_{0}^{\infty} \ \vert \  g_{1/2}^{(-1)}\ \vert^2 \ r^2 d r\ =\ \int_{0}^{\infty} \ \vert \  f_{1/2}^{(-1)}\ \vert^2 \ r^2 d r\ .
\eea
However, the result is quite sensitive to the quark mass, $m$. 
On the contrary, in our case, the magnitude of the small components in the Dirac scalar confining potential is determined by an average {\sl effective mass}, $m_{eff}=\langle m + a r + c \rangle$ (for simplicity, we disregard now the Coulomb potential)~\footnote{We owe this remark to J.-C. Raynal.}. In other words, in our approach, the average $\langle m + ar + c \rangle$, and not just $m$ or $( m + c )$, corresponds to the idea of the constituent mass. This average is large for any reasonable choice of parameters fitting the spectrum ; 
{\it e.g.} for the lowest admissible $( m + c )= -0.2~{\rm GeV}$ and $a\simeq 0.2~{\rm GeV}^2$, $\kappa\in (0.6 \div 0.7)$:
\bea
a\ \langle r \rangle \ \simeq \ 0.2~{\rm GeV}^2 \times 2.45~{\rm GeV}^{-1},
\eea
which gives $\langle m + a r + c \rangle  \simeq 0.3~{\rm GeV}$, therefore, already a
rather large number\footnote{Larger values are obtained at larger $(m + c)$, although $\langle r \rangle$
is smaller: {\it e.g.} for $(m + c) = 0$, we have $m_{eff} = 0.42~{\rm GeV}$, and for $(m + c) = 0.2~{\rm GeV}$, $m_{eff} = 0.55~{\rm GeV}$.}. 
Therefore, the integral over $g_{1/2}^{(-1)}$ ($\lesssim 0.3$), remains always much smaller than over $f_{1/2}^{(-1)}$, for the values of $( m + c )$ compatible with the spectrum. 
In other words {\underline{$\widehat{g}$ remains large}}. 
We reiterate that a smaller $\widehat{g}$ is allowed only when $( m + c )$ gets much below zero, which turns out to fail in the spectrum description. Therefore,  the small values for $\widehat{g}$, like $0.5$ or smaller, are prohibited. 
\par


\subsection{Discussion of other pionic transitions, additivity and saturation of the Adler--Weisberger sum rule} \label{discussion2}

-- Orbital excitations: First, let us notice that our infinite mass limit coupling $h$, is in a quite remarkable agreement with the findings of the QCDSR~\cite{Gatto,Aliev}. On the other hand  the value of $h'/\Lambda_{\chi}$
is compatible with sum rules but somewhat  larger. As for comparison with experiment : 1) As inferred from Tab.~\ref{tab3}, the $D^{**}$ and $B^{**}$ ($j_\ell=3/2$) $2^+$ widths, are reasonably well described by our model. Though, our $D^*_2$ is somewhat too large, which may be explained by large $1/m_Q$ corrections in the
$D$ sector (see in this respect \cite{Mehen}). 2) The predicted large width of order $300~{\rm MeV}$ for $D_0^{*}$ may explain why it is not seen. The experimental total width of the $B^{**}$ bump ($\sim \ 140\ {\rm MeV}$) is smaller than our prediction, $\sim \ 250~MeV$, for the $B_0^{*}$ ; this remains to be understood. 
One interesting feature of this (emission) model is the relative narrowness of the $S$ wave ($0^+$) decays, which were predicted to be much larger (around $1~{\rm GeV}$) by previous quark pair creation models~\cite{Kokoski}. This is actually due to the fact that our model contains an additional factor of $\omega_{\pi}$ in the amplitude.

-- Recently, DELPHI claimed to have detected the very narrow, would be $D^{* '}(2637)$, radial excitation~\cite{Roudeau}. Subsequent search by CLEO~\cite{CLEO98}\footnote{N.B.: In this reference, presented are also the first preliminary results for a broad $J^P=1^+$ charmed state: $m_{D_1} = 2.461^{+.41}_{-.34}\pm .010\pm .032~{\rm GeV}$, $\Gamma = 290^{+101}_{-79}\pm 26\pm 36~{\rm MeV}$.} did not find the trace of such a `peak'. In Ref.~\cite{Melikhov}, the radial excitation interpretation was excluded. Our model is not well suited to discuss radial excitation
decays because of the sensitivity to the nodes of the wave function~\cite{Livre}. With this remark in mind, we give the result
\bea
|X_{D^{* '}}|\ =\ 0.237\, ,\quad \to \quad \Gamma (D^{* '} \to D \pi)\ \simeq \ 70\ {\rm MeV}\ . 
\eea

-- In Ref.~\cite{Pirjol2}, it was concluded that the experimentally measured $\Sigma_c^* \to \Lambda_c + \pi$ width is in agreement with a naive non-relativistic QM estimate, with a small reduction factor. The situation is similar to the one of the $D$ meson, and one expects the validity of the additivity principle: the coupling of light quarks to pion inside $\Sigma_c$ is  the same as in the $D$. This is certainly true for the model with Dirac equation. This observation makes us confident that the value of $\widehat{g}$ is indeed large.

-- For the end, we verify the saturation of the AWSR. We rewrite~(\ref{10}) without radial excitations and multiparticle states:
\bea 
\widehat{g}^2\ +\ {h}^2\ +\ {2\over 3} (\Delta E)^2 {h'^2 \over \Lambda_\chi^2} \ +\ \dots \ =\
 0.368 \ +\ 0.287\ +\ 0.055\ +\ \dots \, ,  
\eea
which means that the first pole and orbital excitations saturate more than $70\%$ the AWSR for $B \pi$ scattering, with the $B^*$ contributing the most. The saturation seems to work with a small number of resonance levels (like in the case of nucleon), in contrast with what would happen if we took the result of QCD sum rules.


\section{Conclusion}
\label{Conclusion}

\begin{itemize}
\item[$\bullet$] Using a Dirac model with a scalar confining potential (which {\em per se} includes the property of additivity of pion couplings), we have obtained a prediction for {$\widehat{g}\ =\ 0.6(1)$}, which is well defined and stable under the variation of the model parameters (chosen in a way compatible with the observed hadron spectrum). This is somewhat different from the previous quark model calculations, which were very sensitive to a (not well constrained) light quark mass parameter. Our model also provides a good description of the decays of orbital excitations, and (presumably) of the charmed baryon decay, $\Sigma_c^* \to \Lambda_c + \pi$; it gives the right $g_A/g_V$, for nucleon~\cite{Luke}. It saturates well the
Adler--Weisberger sum rule with a few low-lying resonances, as observed for the nucleon. The result for $\widehat{g}$ and consequently for $g_{D^{*}D\pi}$ and $g_{B^{*}B\pi}$ is definitely large: $\widehat{g} \geq 0.5$. 

\item[$\bullet$]Of course, the main concern is the complete discrepancy with QCDSR predictions for $\widehat{g}$. It is the first time that one observes such a large discrepancy between the two methods in a situation where both are expected by their proponents, to work reasonably well.

\item[$\bullet$]
Our assumption of a scalar potential (which lacks a QCD basis), the use of an elementary quantum emission model for strong decay to a hadron, and finally the use of the bare quark current for an emission from a light quark (in spite of possibly large renormalization effects), are possible defaults of our approach. Hence, we cannot claim to have compelling and accurate predictions in a quark model. For instance, it is quite possible that one should have to introduce a reduction factor for the quark current $(g_{A})_q$. Still, with all these elements in mind, it would be very surprising if our predictions turn to be totally away from the correct number (as the QCDSR results suggest).

\item[$\bullet$]
Since one can always doubt results of a particular model, we insist on our strongest argument: it would not be easy to escape our (large $\widehat g$) conclusion within any quark model, and to explain the magnitude of the known  pionic couplings of ground state baryons, at the same time. Indeed, if we lessen the strength of $\widehat{g}$, then a similar suppression in the pionic couplings in other ground state hadrons (due to approximate additivity for light quarks pion couplings, expected at least to work well for charmed hadrons) would be difficult to escape. Now, such a suppression is not actually observed. It is neither possible to introduce a small $(g_{A})_q$ in order to reconcile the quark model with the small result of sum rules, because it will be still more unavoidable to end with such a suppression in other pionic couplings, which is
not observed. 

\item[$\bullet$] On the side of QCDSR, certainly the convergence of the various calculations towards a small value of $\widehat{g}$ is very impressive. 
However, one should:
\begin{itemize}
\item[{\sf a)}] have a conceptual understanding of why the
result is so small, as compared to the naive non-relativistic QM 
($\sim 1/5$ suppression factor!), while the latter works reasonably for the analogous $g_{\rho\pi\pi}$, and $g_{K^*K\pi}$ couplings ({\it e.g.} recent result of Ref.~\cite{Savci});

\item[{\sf b)}] explain why the heavy baryon widths 
apparently do not exhibit any particularly strong suppression with respect to a naive non-relativistic prediction;

\item[{\sf c)}] understand how the
AWSR is saturated, with the ground state
contribution being so small.
\end{itemize}
\end{itemize}

\acknowledgments
We would like to thank J.-C.~Raynal for useful discussions,
as well as A.~Khodjamirian for useful communication.
D.B. acknowledges the partial support of~{\em ``La~Fondation~des~Treilles''}.
\newpage

\end{document}